\documentclass[useAMS,usenatbib]{mn2e}
\usepackage{graphicx}

\title[Constraints on Exotic Matter for An Emergent Universe]{Constraints on Exotic Matter Needed for An Emergent Universe
}
\author[B. C. Paul, P. Thakur and S. Ghose]
  { B. C. Paul,$^{1,}$$^3$ \thanks{ Electronic mail : bcpaul@iucaa.ernet.in }
  P. Thakur,$^{2,}$$^3$ \thanks{ Electronic mail : prasenjit \textunderscore thakur1 @yahoo.co.in}
  S. Ghose$^{1,}$$^3$ \thanks{ Electronic mail : souvik@bose.res.in} 
   \\
  $^1$Physics Department, North Bengal University \\ 
      Dist. : Darjeeling, Pin : 734 013, West Bengal, India 
\\
  $^2$Physics Department, Alipurduar College\\
      Dist. : Jalpaiguri, Pin : 736122, West Bengal, India
 \\
 $^3$ IUCAA Reference Centre, Physics Department \\
 North Bengal University \\}
\date{}
\begin{document}


\pagerange{\pageref{firstpage}--\pageref{lastpage}} \pubyear{2008}

\maketitle

\label{firstpage}
\begin{abstract}

We study  a composition of normal and exotic matter which is required for a flat Emergent Universe scenario permitted by the equation of state (EOS)($p=A\rho-B\rho^{\frac{1}{2}}$) and predict the  range of the permissible values for the parameters $A$ and $B$ to explore a physically viable cosmological model. The permitted values of the parameters are determined taking into account the $H(z)-z$ data  obtained from observations, a model independent BAO peak parameter and CMB shift parameter (WMAP7 data). It is found that although $A$ can be very close to zero, most of the observations favours a small and negative $A$. As a consequence, the effective Equation of State parameter for this class of Emergent Universe solutions remains negative always. We also compared the magnitude ($\mu (z)$) vrs. redshift($z$) curve obtained in the model with that obtained from the union compilation data. According to our analysis the class of Emergent Universe solutions considered here is not ruled out by the observations.   

\end{abstract}

\begin{keywords}
Emergent universe, Exotic matter.
\end{keywords}

\section{Introduction}

It is well known from the recent observations that the standard Big Bang model of cosmology fails to describe the present accelerating phase of the universe. The model is also pleagued by a time like singularity in the past. Accelerating phase of the universe can, however, be incorporated in a number of ways. A number of models e.g., models with modified theory of gravity \citep{b19}, models with unusual matters like Chaplygin gas \citep{b3,b4}, scalar and tachyon fields \citep{b18} are taken into account to accommodate present phase of acceleration. There are other models based on mostly non-equilibrium thermodynamics and Boltzmann formulation which do not require dark energy \citep{b15,b16,b17}.   Only very recently some other models appeared in the literature which discusses cold dark matter (CDM) and CDM interactions as alternative to the $\Lambda$CDM model \citep {n1,n2}. However, exploring singularity free cosmological models is an interesting area in cosmology and Emergent Universe scenario (EU) is one of the well known choices. A number of literature appeared which discussed EU model as it was free from initial singularity and the size of the universe was large enough so that quantum gravity effects were not important \citep{b8,b7}. These models evolve from a static phase in the infinite past into an inflationary phase. The idea is in conformity with Lemaitre-Eddington concepts from early days of modern cosmology. If developed in a consistent manner an emergent universe model is capable of solving  some of the well known conceptual problems  not understood in the Big-Bang model. A model of an ever-existing universe, which eventually enters into the standard Big Bang epoch at some stage and consistent with features known to us today is worth considering. Recently an interesting EU model has been proposed by \citet{b12} which requires some exotic matter in addition to normal matter as cosmic fluid. The model has been explored in a flat universe as such universe is supported by recent observations. Subsequently the EU model was taken up to examine the suitability of implementing it in the context of various theories \citep{b1,b5,b2,b13}.  The EOS needed for the model proposed by \citet{b12} is given by
\begin{equation}
p=A\rho-B\rho^\frac{1}{2},
\end{equation}
where $A$ and $B$ are unknown parameters of the theory. We note that different values of $A$ and $B$ pick up different composition of matters which may lead to an EU model. A similar EOS was considered in the literature as a double component dark energy model \citep{w2} where the model parameters are constrained from Type Ia supernova data. The EOS considered by them is basically a special form of a more general EOS, $p=A\rho -B\rho^{\alpha}$; which may be seen as a realization of Chaplygin gas (with $\alpha <0$) \citep{b3,b4}. It may be mentioned here that Chaplygin gas models were introduced in cosmology as an interpolation between a matter dominated era and a de Sitter phase. Later a modified model of Chaplygin gas was proposed \citep{b10} to describe cosmological evolution. For example models like Modified Chaplygin gas serves well as an interpolation between radiative era and $\Lambda$CDM era. \citet{w2} showed that even with $\alpha>0$ such interpolation is permissible and an EOS like one considered by \citet{b12} may be considered as a phenomenological realization of string specific configuration.  The model proposed by \citet{b12} developed the EU scenario in a very general way. Once one considers an EOS given by eq. (1), a class of EU solutions is permitted for $B>0$. The authors however showed that the above EU solution are not permitted with a minimally coupled scalar field. Also, it was found that the EU scenario automatically admitted a composition of three kinds of matter energy density in the universe all having their own way of evolution. This is certainly an interesting issue keeping in mind that the model has the provision for a large class of possible dark energy and dark matter candidate. It is thus worth to investigate the viability of such an EU model with the recent observational data. Nevertheless we intend to explore the allowed range of values of  the parameter $A$, for $B>0$ for a viable cosmological scenario permitted by observations. To determine the range of values for $A$ and $B$ for a viable cosmological model permitted by observations, we adopt here two independent techniques:
(i)  Applying $\chi^2$ minimization technique on $H(z)$ vs. $z$ data \citep{b14}. Here we use $9$ data points given in Table 1, (ii) using joint analysis of $H(z)$ vs. $z$ data and a  model independent BAO peak parameter and (iii) using joint analysis of $H(z)$ vs. $z$ data, BAO peak parameter and CMB shift parameter. We explore here the suitability  of the model with the help of supernovae data (union compilation data) also. The plan of the paper is as follows : In sec.2 relevant field equations are obtained from Einstein field equation. In sec.3 the constraints on model parameters are determined  from $H(z)$ vs. $z$ data. Subsequently in sec. 4 and sec. 5 we obtain more stringent constraints on model parameters in accordence with the joint analysis with  model independent BAO peak parameter and CMB shift parameter. In sec.5, we draw the $\mu (z) -z$ curve for our model to compare with that drawn using union compilation data \citep{b9}. Finally, in sec. 6 we summarize our results and briefly discuss the results.\\
 \begin{table}
  \begin{minipage}{140mm}

  \caption{$H(z) vs. z$ data}
  \begin{tabular}{l|c|r}
  \hline
  {\it z Data} & $H(z)$ & $\sigma$ \\
  \hline
   0.00 & 73  & $ \pm $ 8.0	 \\
   0.10 & 69  & $ \pm $ 12.0 \\
   0.17 & 83  & $ \pm $ 8.0 \\
   0.27 & 77  & $ \pm $ 14.0 \\
   0.40 & 95  & $ \pm $ 17.4 \\
   0.48 & 90  & $ \pm $ 60.0 \\
   0.88 & 97  & $ \pm $ 40.4 \\
   0.90 & 117 & $ \pm $ 23.0 \\
   1.30 & 168 & $ \pm $ 17.4 \\
   1.43 & 177 & $ \pm $ 18.2 \\
   1.53 & 140 & $ \pm $ 14.0 \\
   1.75 & 202 & $ \pm $ 40.4 \\

\hline
\end{tabular}
\end{minipage}
\end{table}

\section{ Field Equations}

We consider Friedmann-Robertson-Walker(FRW) metric (c=1), given by :

\begin{figure}
\includegraphics[width=240pt,height=200pt]{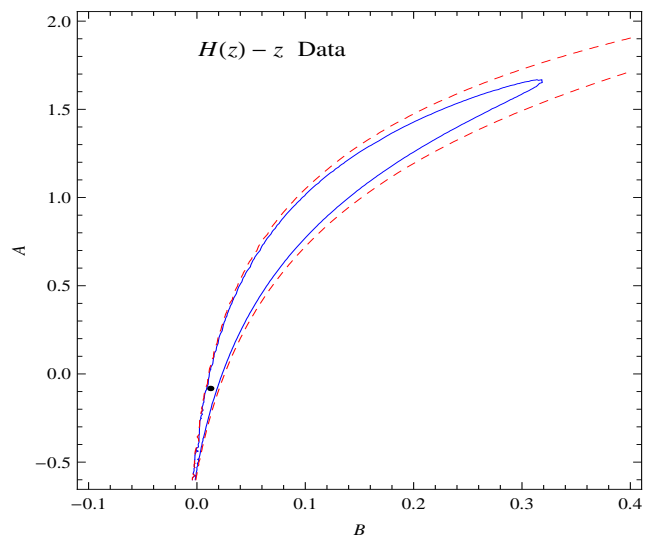}
\caption{(Colour Online)Constraints from $H(z) \: vs. \: z$ for $K=0.0100$.  (Solid) $99 \%$ and $95 \%$  (Dashing) contours. The best fit point is shown ($0.0122, -0.0823$).}
\end{figure}

\begin{equation}
ds^{2} = - dt^{2} + a^{2}(t) \left[ \frac{dr^{2}}{1- k r^2} + r^2 ( d\theta^{2} + sin^{2} \theta \;
d  \phi^{2} ) \right]
\end{equation}
where  $k=0,+1(-1)$ is the curvature parameter in the spatial section representing flat, closed (open) universe and $a(t)$ is the scale factor of the universe and $r,\theta,\phi$ are the dimensionless comoving co-ordinates.
The energy conservation equation is given by 
\begin{equation}
\frac{d\rho}{dt} + 3 H (\rho + p) = 0 ,
\end{equation}
where $p$, $\rho$ and $H$ are respectively pressure, energy density, Hubble parameter.
We express Hubble parameter ($H$) in terms of redshift parameter ($z$) which is given by
\begin{equation}
H(z)=-\frac{1}{1+z}\frac{dz}{dt}.
\end{equation}
Since components of matter (baryon) and Dark energy (exotic matter) are conserved separately, we use energy conservation equation together with EOS given by (1) to determine the expression for the energy density. Consequently eq. (3) yields:

\begin{equation}
\rho_{emu}= \left[\frac{B}{1+A}+\frac{1}{A+1}\frac{K}{a^{\frac{3(A+1)}{2}}}\right]^{2},
\end{equation}
where  $K$ is an integration constant which is required to be positive definite. From eq. (5) it is evident that the energy density is composed of three different terms, where a constant term ($(\frac{B}{1+A})^2$)  may be identified with a cosmological constant and  the other two terms are identified with two kinds of fluids determined by the parameters $A$. For simplicity, eq. (5) can be rewritten as,
\begin{equation}
\rho_{emu}=\rho_{em_{0}}\left[A_s+\frac{1-A_s}{a^{3(A+1)/2}}\right]^{2}
\end{equation}
where  $A_s=\frac{B}{1+A}\frac{1}{\rho_{em_0}^{\frac{1}{2}}}$ and $\frac{K}{A+1}=\rho_{em_0}^{\frac{1}{2}}-\frac{B}{A+1}.$\\
Using the Friedmann  equation we express $H$ in terms of redshift parameter ($z$) for the model,  which is given by 

\begin{eqnarray}
H(z) & = & H_0 [ \Omega_{b_{0}}(1+z)^3+  \nonumber \\
     &   & (1-\Omega_{b_{0}})  [\frac{B+K(1+z)^{\frac{3(A+1)}{2}}}{B+K}]^{2}]^{\frac{1}{2}},
\end{eqnarray} with $\Omega = \Omega_{b_{0}}+\Omega_{em_{0}}=1$, where $\Omega$ is composed of baryon and exotic  fluids. $\Omega_{b_0}$ represents baryon energy density and $\Omega_{em_0}$ represents the exotic fluid density. Here we consider $\Omega_{b_0}=0.04$ \citep{n5}.
\section{H(z)-z Data as a constraining tool}

The EU considered here is implemented in a flat universe, consequently we consider  baryonic matter and the exotic matter in a flat Friedmann universe to the allowed range of the parameters. The Hubble parameter given by (7) is a function of a number of variables, consequently we can re-write  eq. (7) as :

\begin{equation}
H^{2}(H_{0},A,B,K,z)=H^{2}_{0}E^{2}(A,B,K,z),
\end{equation}
where
\begin{eqnarray}
E(A,B,K,z) & = & [\Omega_{b_{0}}(1+z)^3
+ \nonumber \\
           &   &   (1-\Omega_{b_{0}})[\frac{B+K(1+z)^{\frac{3(A+1)}{2}}}{B+K}]^{2}]^{\frac{1}{2}}.
\end{eqnarray} 
For a given value of $K$, the best fit values for the unknown parameters of the model, namely $A$ and $B$ are determined by minimizing a $\chi^2_{H-z}$ function which is given below.
\begin{equation}
\chi^{2}_{H-z}(H_{0},A,B,K,z)=\sum\frac{(H(H_{0},A,B,K,z)-H_{obs}(z))^2}{\sigma^{2}_{z}}
\end{equation} where $H_{obs}(z)$ is the observed Hubble parameter at redshift $z$ and $\sigma_z$ is the error associated with that particular observation. Since we are interested in determining the model parameters, $H_{0}$ is not important for our analysis. So we marginalize over $H_{0}$ to get the probability distribution 
function in terms of  $A,B,K$ only, which is given by
\begin{equation}
L(A,B,K)=\int dH_{0}P(H_{0})\exp \left(\frac{-\chi^{2}_{H-z}(H_{0},A,B,K,z)}{2} \right),
\end{equation}
where $P(H_{0})$ is the prior distribution function
for the present Hubble constant. Here we  consider
Gaussian priors, $H_{0}=72 \pm 8$ . One can minimize $\chi^2$ by maximizing the function $L(A,B,K)$. We fix $K$ at the best fitted value and contours in $A$-$B$ plane are drawn. However, fixing of $K$ is allowed as we are interested to obtain range of $A$ and $B$ which is related to the EOS given by eq. (1). $K$ enters in the theory as an integration constant which is always positive ($K>0$). In fig. 1 we draw $99 \%$ and $95 \%$  contours on $A$-$B$ plane. We see that within $99 \%$ confidence $-0.5949 \leq A \leq 1.663 $ and $-0.0022 \leq B \leq 0.3189$. Of course theoretically we must have $B \geq 0$ to obtain an EU solution for the model we are considering here.

\begin{figure}
\includegraphics[width=240pt,height=200pt]{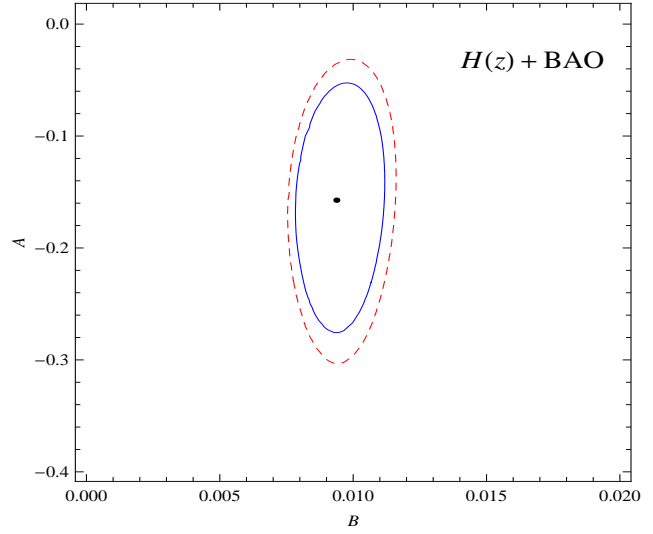}
\caption{(Colour Online)Constraints from joint analysis with $H(z)$-$z$ data BAO peak parameter for $K=0.0101$. $99 \%$ (Solid) and $95 \%$ (Dashing) confidence level are shown in the figure along with the best fit value(0.0094,-0.1573)}
\end{figure}

\section{Joint analysis of $H(z)$-$z$ with BAO peak parameter}

In this section we use the technique adopted by \citet{b6} to explore the parameter $\mathcal {A}$ (which is determined by $A$). The value of $\mathcal {A}$ is independent of the cosmological models, which for a flat universe can be expressed as :

\begin{equation}
\mathcal {A} =\frac{\sqrt{\Omega_{m}}}{E(z_{1})^{1/3}}\left(\frac{\int ^{z_1}_0 \frac{dz}{E(z)}}{z_{1}}\right)^{2/3}
\end{equation}
where $\Omega_{m}=\Omega_{b}+(1-\Omega_{b})(1-B/(K+B))^2$ and $z_{1}=0.35$ and $\mathcal {A} =0.469 \pm 0.017$.
 We define $\chi ^2_{BAO} = \frac{(\mathcal {A} - 0.469)^2}{(0.017)^2}$, and for joint analysis we consider $\chi ^2_{joint}= \chi ^2_{H-z} + \chi ^2_{BAO}$.

The above joint analysis scheme with BAO sets  new constraints on $A$ and $B$, again which are upto $95 \%$ confidence level $-0.3053 \leq A \leq -0.0306 $ and $0.0077 \leq B \leq 0.0116 $ and upto $99 \%$ confidence level $-0.2757 \leq A \leq -0.0500 $ and $0.0078 \leq B \leq 0.0114 $.

\begin{figure}
\includegraphics[width=240pt,height=200pt]{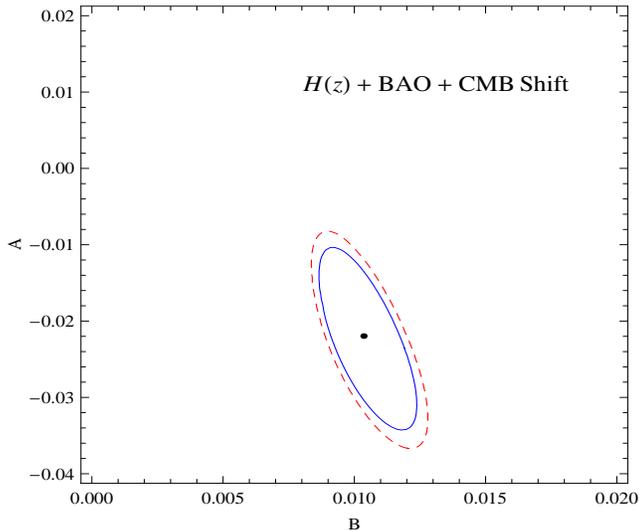}
\caption{(Colour Online)Constraints from joint analysis of $H(z)$-$z$ data, BAO peak parameter and CMB shift parameter for $K=0.0101$. $99 \%$ (Solid) and $95 \%$ (Dashing) confidence level are shown in the figure along with the best fit value(0.0102,-0.0176)}
\end{figure}
\section{ Joint analysis with $H(z)$-$z$ , BAO peak parameter and CMB shift parameter ($\mathcal {R}$)}

CMB shift parameter ($\mathcal {R}$) is given by

\begin{equation}
\mathcal{R}=\sqrt{\Omega_{m}}\int ^{z_{ls}}_{0} \frac{dz'}{H(z')/H_{0}}
\end{equation}
where $z_{ls}$ is the $z$ at the surface of last scattering. The WMAP7 data gives us $\mathcal{R}=1.726 \pm 0.018$ at $z=1091.3$ \citep{w1}. Thus we consider $\chi^2_{CMB}=\frac{(\mathcal{R}-1.726)^2}{(0.018)^2}$, with $\chi ^2_{Tot} = \chi ^2_{H-z}+\chi ^2_{BAO}+\chi ^2_{CMB}$ which impose additional constraints on the model parameters. The statistical analysis with $\chi ^2_{Tot}$ further tightens up the bounds on $A$ and $B$. In fig. 3, $95\%$ and $99\%$ contours are plotted on $A$-$B$ plane. We determine constraints from this analysis: within $95 \%$ confidence level $-0.037 \leq A \leq 0.008$  and $0.008 \leq B \leq 0.013 $. However, within $99 \%$ confidence limit we get $ -0.034 \leq A \leq -0.0114$ and $0.012 \leq B \leq 0.009$. The best fit value obtained here is given by  $A=-0.0219$ and $B=0.0103$. Finally we draw a magnitude ($\mu(z)$) vs. redshift ($z$) curve for our model with the best fit values of $A$, $B$ and $K$ and also show the same curve drawn from union compilation data for SNeIa \citep{b9} in fig. 4. 

\section{Discussions}

\begin{figure}
\includegraphics[width=240pt,height=200pt]{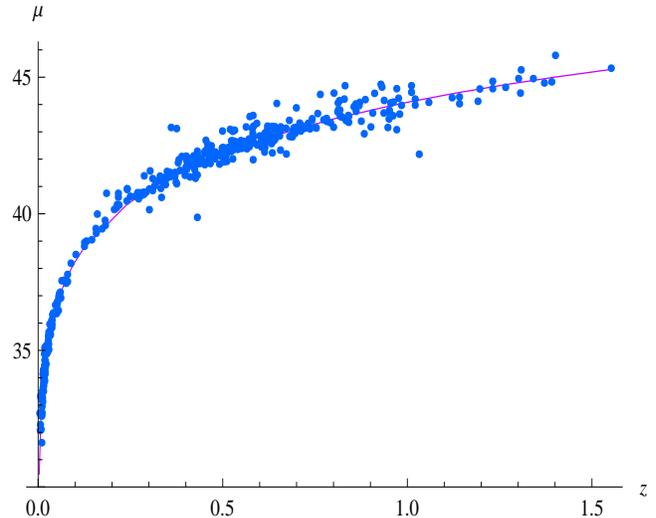}
\caption{(Colour Online) $\mu(z) \: vs. \: z$ curve comparism with supernovae data}
\end{figure}

\begin{figure}
\includegraphics[width=240pt,height=200pt]{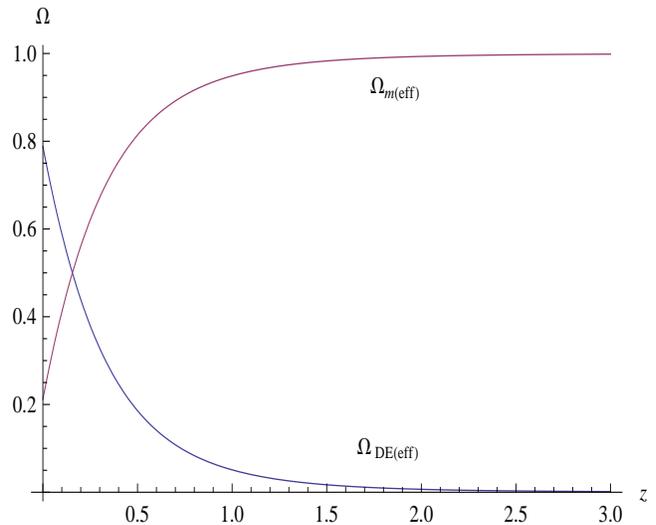}
\caption{(Colour Online) Plot of density parameter ($\Omega$) for effective dark energy and effective matter content of Universe.}
\end{figure}
Theoretically the class of EU solutions considered here can be realized for a composition of different kinds of matter \citep{b12} depending on the model parameters $A$ and $B$. In this paper we determine allowed ranges for the model parameters (particularly those involved with the EOS i.e., $A$ and $B$). Considering a EOS required for a flat emergent universe we determine the constraints on the parameters using data available from cosmological observations. We note from the analysis that the EOS that permits a class of EU solution, considered here, should contain exotic matter ($A<0$, $B>0$). This is certainly not ruled out by the theory itself. It may be pointed out here that $H(z)$-$z$ data puts a bound on the model parameters which is further investigated in the light of the other observational data such as BAO peak parameter value and CMB shift parameter. In the above we estimate model parameters ($A$, $B$ and $K$). Most important point to be noted here that the later observations do not permit a positive value for the parameter $A$. Only small negative values seem to be allowed. Although positive $A$ values are permitted when we consider $H(z)$-$z$ data only but the best fit value is found to be negative. However the possibility that $A \approx 0$ can not be entirely ruled out since our analysis permits values of $A$ which are even very close to zero ($A=0$) and the model may be realized in the presence of dust and dark energy. We also study the evolution of various cosmological parameters of the model. For example density parameter is on such important parameter. We plot density parameter for effective dark energy and effective matter content of the universe with the redshift in fig. 5. We note that almost $80\%$ of the present matter-energy content is effective dark energy and baryonic and nonbaryonic matter constitutes the remaining part. So, as far as present budget of Dark Energy and Dark matter is concerned, EU differes very little from $\Lambda$CDM model. However, as mentioned earlier, the class of EU solution considered here has provision for different composition of matter-energy in universe depending on the values of the parametes $A$ and $B$. It can also accommodate a cosmological constant in a special case. The effective equation of state ($\omega_{eff}$) for EU remains negative always which is evident from fig. 6. The solid line in fig. 6 corresponds to the curve drawn using best fitted values. Dash and dotted curves are drawn with typical model parameters values within 95$\%$ and 99$\%$ confidence leves respectively. The transition of the universe from a deceleration phase to an accelerating phase in recent past is depicted from the curve of  deceleration parameter against redshift plotted in fig. 7. The solid curve describes the one drawn with best fitted values and dotted and dash  curves represent curves drawn with values within 99\% and 95\% confidence level respectively.  We conclude that the class of EU solutions considered here is not ruled out by the observations. However, this class of EU solutions admits different composition of matter-energies in the universe and the nature of composition depends on the value of parameter $A$ in particular. The observations do in fact severely constrain the nature of allowed composition of matter-energy by constraining the range of the values of the parameters for a physically viable model.

\begin{figure}
\includegraphics[width=240pt,height=200pt]{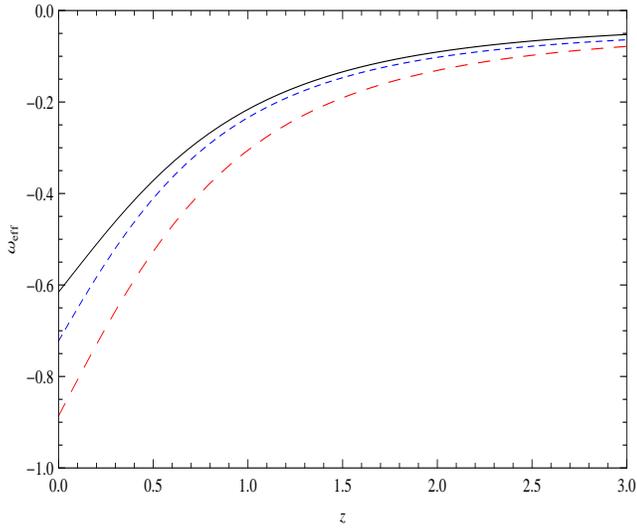}
\caption{(Colour Online) Effective EOS parameter for EU ( $\omega_{eff}$ ) is plotted with redshift. Plot with the best fitted values of model parameters $k=0.0102$, $A=-0.0176$ and $B=0.0103$ (Solid). Plot with values within 95\% confidence (Dashing) and within 99\%  (Dotted).}
\end{figure}

\begin{figure}
\includegraphics[width=240pt,height=200pt]{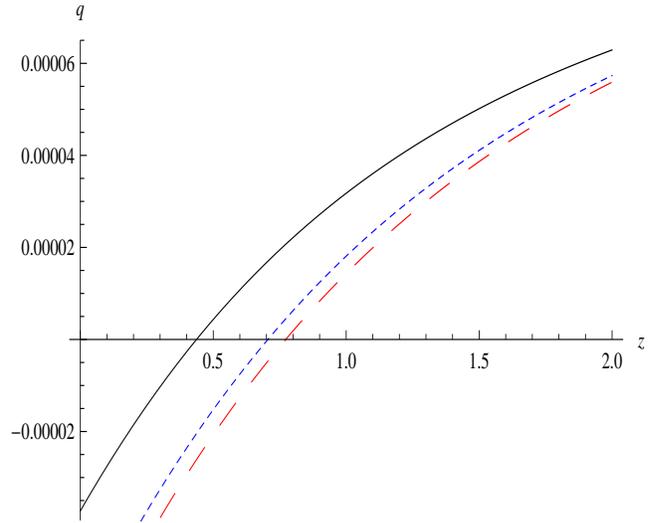}
\caption{(Colour Online)Deceleration parameter ($q$) vs. redshift ($z$). Plot with the best fitted values of model parameters(Solid), with values within 95\% confidence (Dashing) and within 99\%  (Dotted).}
\end{figure}

\section*{Acknowledgments}
BCP, PT and SG would like to thank {\it IUCAA Reference Centre}, Physics Department, N.B.U for extending the facilities of research work. SG would like to thank University of North Bengal for awarding Senior Research Fellowship. BCP would like to thank University Grants Commission, New Delhi for Minor Research Project (No. 36-365/2008 SR). BCP would like to thank IUCAA, Pune for support under visiting associateship program where a part of the work is done. SG would like to thank SNBNCBS, Kolkata for support under visiting student associateship program. Authors would like to thank the referee for his suggesstions which helped to improve the work.

\end{document}